# Thermal capacities and polytropic processes

Joaquim Anacleto

Departamento de Física da Escola de Ciências e Tecnologia da Universidade de Trás-os-Montes e Alto Douro, Quinta de Prados, 5000-801 Vila Real, Portugal

IFIMUP-IN e Departamento de Física e Astronomia, Faculdade de Ciências, Universidade do Porto, R. do Campo Alegre s/n, 4169-007 Porto, Portugal

e-mail: anacleto@utad.pt

ORCID: https://orcid.org/0000-0002-0299-0146

**Abstract**

Herein, we analyse polytropic processes for an ideal gas within the wider concept of thermal capacity. To answer the question of whether the thermal capacity is a process, path, or state function, we argue that it should be tentatively set as a path function and if it remains constant along the path, the corresponding process is polytropic. Of all the paths, there are only two – at constant volume and constant pressure – for which the thermal capacities – $C_V$ and $C_P$ – are state functions, i.e., system properties. The discussions herein are valuable both scientifically and instructively because they shed light on issues in undergraduate curricula that are not addressed in sufficient detail in physics textbooks, not even in the most advanced ones.

## 1. Thermal capacities: process or path functions?

A typical experimental task in teaching thermodynamics is determining the thermal capacity of a system, for example, a given amount of water. One possible approach is to heat the system using a flame and measure its temperature rise. The infinitesimal heat $\delta Q$ required for an infinitesimal temperature change $dT$ is related to the *thermal capacity C* as [1]

$$C = \frac{\delta Q}{dT}. \qquad (1)$$

Despite the conceptual simplicity of this experiment, it is difficult to measure the heat entering the system from the flame. Therefore, from an experimental perspective, instead of using a flame, it is preferable to heat the system using an electrical resistance $r$ placed inside it and becoming a part of it; the resistance is connected to a battery with an electromotive force $\varepsilon_e$, which is placed in the surroundings. In this setup, the energy supplied to the system *is not* *heat* but *dissipative* (electrical) *work*, $W_D$, whose differential is given by $\delta W_D = \left(\varepsilon_e^2/r\right) dt$ (see Appendix), where $dt$ is the infinitesimal heating time. The thermal capacity is related to $\delta W_D$ and $dT$ as

$$C = \frac{\delta W_D}{dT}. \qquad (2)$$

Comparing (1) and (2), the effects of $\delta Q$ and $\delta W_D$ on the system are indistinguishable [2]. However, although the sequence of system states follows the *same path* when represented in a diagram, the increase in temperature via heat or dissipative work are two *distinct processes* because while heat leads to the variation of the entropy of both the system and surroundings, dissipative work leads only to the variation of the entropy of the system.

*Path* and *process* are different concepts. A path is defined as the sequence of system states described by a mathematical relation involving *only* the system variables. In contrast, a process is an interaction between a system and its surroundings, described not only by the system variables but also by those of the surroundings.

Assuming (1) is the definition of *C*, the thermal capacity is a *process function* but not a *path function* because a system following a given path may have different values for *C* depending on whether $\delta Q$ or $\delta W_D$, or both, exist. The same is true if (2) is considered as the definition of *C*. Therefore, neither (1) nor (2) alone are adequate definitions as they are inconsistent with the experimental determination of *C*, yielding different values based on the process, i.e., depending on whether heat, dissipative work, or both are used in the heating process.

The above considerations lead us to merge the two (inadequate) definitions (1) and (2) into a single definition that is consistent not only with the experimental determination of *C* but also with the fact that both $\delta Q$ and $\delta W_D$ produce indistinguishable effects on the system (in particular, the same temperature change) [2]. Accordingly, we propose the relation

$$C = \frac{\delta Q + \delta W_D}{dT}. \qquad (3)$$

Definition (3), which is virtually absent in the existing literature, establishes *C* as a *path function*, which, as discussed earlier, is different from a *process function*. This definition is more suitable than the previous ones because it does not depend on surroundings variables but instead relies solely on the evolution of the system during the process. The next question we consider is: can *C* also be a *state function*? If so, in which cases?

## 2. Thermal capacities as state functions

To confirm that the definition of *C* in (3) is indeed a path function, we consider an equivalent expression where only the system variables appear explicitly. As shown in the Appendix, from the fundamental equation that describes a process [3], we have

$$\delta Q + \delta W_D = T dS, \qquad (4)$$

where *S* is the system entropy. Inserted (4) in (3), we obtain

$$C = \frac{\delta Q + \delta W_D}{dT} = T \frac{dS}{dT}. \qquad (5)$$

Thus, at each point on a *S–T* path, *C* is determined by the temperature and slope. In another representation, for example, a *V–T* path, *S* can be considered as a function of *T* and *V*, i.e., $S = S(T,V)$; thus, d*S* in (5) can be replaced by

$$dS = \left(\frac{\partial S}{\partial T}\right)_V dT + \left(\frac{\partial S}{\partial V}\right)_T dV. \qquad (6)$$

Consequently,

$$C = T\left(\frac{\partial S}{\partial T}\right)_V + T\left(\frac{\partial S}{\partial V}\right)_T \frac{dV}{dT}. \qquad (7)$$

Using the *thermal expansion* and *compressibility* coefficients, $\alpha$ and $\kappa_T$, respectively, which are defined as

$$\alpha = \frac{1}{V}\left(\frac{\partial V}{\partial T}\right)_P \qquad (8)$$

and

$$\kappa_T = -\frac{1}{V}\left(\frac{\partial V}{\partial P}\right)_T, \qquad (9)$$

we obtain $(\partial S/\partial V)_T = \alpha/\kappa_T$ [3]. Therefore, (7) can be rewritten as

$$C = T\left(\frac{\partial S}{\partial T}\right)_V + T\frac{\alpha}{\kappa_T}\frac{dV}{dT}. \qquad (10)$$

Both (5) and (10) indicate that *C*, as defined in (3), is a *path function* and is expressed only in terms of system variables. This is in contrast to (1) and (2), wherein $\delta Q = -T_e dS_e$ and $\delta W_D = \left(\varepsilon_e^2/r\right)dt$, with $T_e$ and $S_e$ being the temperature and entropy of the surroundings, respectively, and $\varepsilon_e$ being the battery electromotive force, which is also a surroundings variable.

From (5), for isentropic processes, $dS = 0$ and $C = 0$, and for isothermal processes, $dT = 0$ and $C = \infty$. Among all other possible paths, there are two in which the thermal capacity becomes a *state function*, i.e., a *system property*. Using (10), one of these paths is defined by $dV = 0$ (an isometric process), where $C$ becomes the *heat capacity at constant volume* $C_V$:

$$C_V = T \left( \frac{\partial S}{\partial T} \right)_V. \tag{11}$$

The other path is defined by $dP = 0$ (an isobaric process), where $C$ becomes the *heat capacity at constant pressure* $C_P$:

$$C_P = T \left( \frac{\partial S}{\partial T} \right)_V + T \frac{\alpha}{\kappa_T} \left( \frac{\partial V}{\partial T} \right)_P. \tag{12}$$

Considering (8) and (11), (12) becomes

$$C_P = C_V + TV \frac{\alpha^2}{\kappa_T}. \tag{13}$$

To highlight the ability of the thermodynamics formalism, we consider the path $dS = 0$ (an isentropic process) and use (5), (10) and (11) to obtain an equivalent expression for $C_V$:

$$C_V = -T \frac{\alpha}{\kappa_T} \left( \frac{\partial V}{\partial T} \right)_S. \tag{14}$$

This illustrates the importance of $C_V$ and $C_P$: they are *state functions* constituting important system properties, whereas all the other thermal capacities, given by (5) and (10), are *path functions*. Thus, $C_V$ and $C_P$ depend *only* on the system state, and the indices $V$, $P$ and $S$ on the right-hand sides of (11), (12) and (14) do not refer to specific paths but rather denote the variables to be held constant during partial derivation. This is an important point that is repeatedly misunderstood.

Simple systems (also called *PVT* systems) are characterised by several properties, only three of which are independent. The easiest properties to measure are $C_P$, $\alpha$ and $\kappa_T$; consequently, these are selected as the fundamental properties [4, 5] to which all the others can be related; for example, $C_V$ can be obtained from (13).

### 3. Polytropic processes of an ideal gas

For liquids and solids, $\alpha^2/\kappa_T \approx 0$, which implies that $C_V \approx C_P$, based on (13). However, this is not true for gases, for which $C_V$ and $C_P$ differ significantly. Considering an ideal gas, the

state equation is

$$PV = nRT, \tag{15}$$

where $n$ is the amount of gas (in mol) and $R = 8.314$ Jmol$^{-1}$K$^{-1}$ is the universal gas constant.

Applying (15), (8) and (9) for an ideal gas turn into

$$\alpha = \frac{1}{T}, \tag{16}$$

and

$$\kappa_T = \frac{1}{P}. \tag{17}$$

Inserting these coefficients into (10) and using (11), we obtain

$$C = C_V + P\frac{dV}{dT}. \tag{18}$$

Adopting the definition in [6], a *polytropic process* of an ideal gas is one wherein the thermal capacity remains constant throughout the entire process. Therefore, taking $C_\beta$ as a constant, where $\beta$ is the polytropic index, (18) can be expressed as

$$C_V + P\frac{dV}{dT} = C_\beta. \tag{19}$$

Using the state equation (15) we obtain

$$\left(\frac{nR}{C_\beta - C_V} - 1\right)\frac{dV}{V} = \frac{dP}{P}, \tag{20}$$

which, when integrated, yields

$$PV^\beta = K, \tag{21}$$

where $K$ is another constant and $\beta$ is given by

$$\beta = 1 - \frac{nR}{C_\beta - C_V}. \tag{22}$$

Typically (21) is considered the definition of a polytropic process [7]; however, as argued in [6], we consider it to be a result rather than a definition. In a $P$–$V$ diagram, a polytropic process with an index $\beta$ follows the path defined by (21) and, by (22), its thermal capacity $C_\beta$ is

$$C_\beta = C_V + \frac{nR}{1-\beta}. \tag{23}$$

Figure 1 shows the graph of $C_\beta$ as a function of the polytropic index $\beta$ along with the thermal capacities for isobaric $(\beta = 0, C_0 = C_P)$, isothermal $(\beta = 1, C_1 = \infty)$, isentropic $\left(\beta = \gamma = C_P/C_V, \; C_\gamma = 0\right)$ and isometric $\left(\beta = \infty, \; C_\infty = C_V\right)$ processes. By analysing the graph, we conclude that if $1 < \beta < \gamma$, then $C_\beta < 0$; these are processes for which, by (5), the changes in temperature and entropy of the system have opposite signs [6].

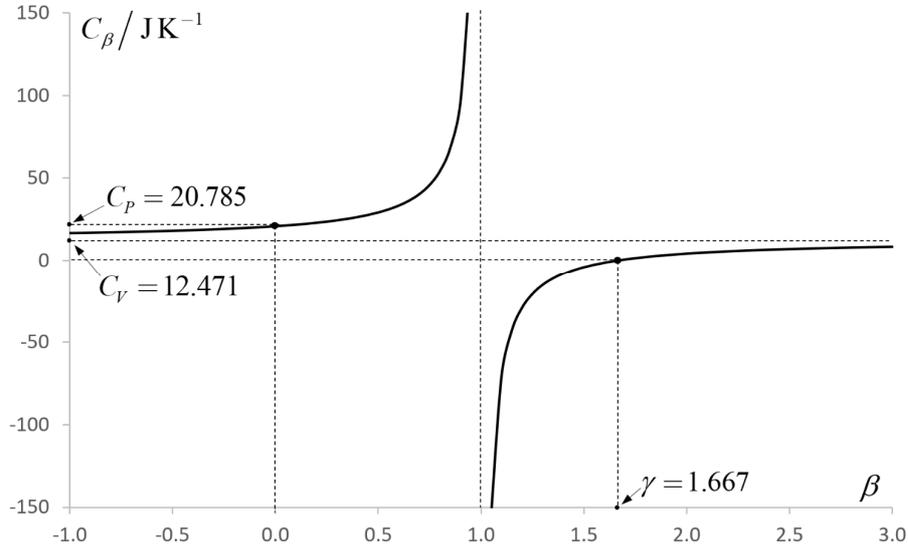

**Figure 1.** Thermal capacities $C_\beta$ for polytropic processes for one mole of a monatomic ideal gas. $\beta$ is the polytropic index and $\gamma = C_P/C_V$ is the adiabatic coefficient. $C_0 = C_P$ (isobaric); $C_1 = \infty$ (isothermal); $C_\gamma = 0$ (isentropic); $C_\infty = C_V$ (isometric).

If the system is not an ideal gas, it is difficult to obtain an analytical expression for a polytropic process because not only is $C_V$ not constant but also (15)–(17) are no longer valid. In this case, the curve describing a polytropic process must be determined by solving (10) numerically while imposing a constant value for $C$, which requires that $C_V$, $\alpha$ and $\kappa_T$ are known throughout the process.

## 4. Conclusions

Herein, we analysed the concept of thermal capacity and ascertained the situations in which it is a process, path, or state function. This distinction between process, path, and state is a novel approach that clarifies certain misunderstandings related to this topic, namely, regarding the essential role of dissipative work in conjunction with heat. In general, the thermal capacity is a path function except for two special cases – constant volume and constant pressure processes, wherein their thermal capacities are also state functions, i.e., system properties. Finally, we discussed the polytropic processes of an ideal gas, from a new perspective within the thermal capacity formalism.

**Appendix**

For reference, (4) is derived here. The application of this formalism to the introductory problem is also discussed. Further details can be found in references [2, 3].

A thermodynamic system is characterised by the macroscopic, or state, variables. These comprise temperature $T$, entropy $S$, and mechanical variables, the specifics of which depend on the problem under study and are typically referred to as generalised forces $Y$ and generalised displacements $X$ [1]. The conservation of energy principle, in conjunction with the *fundamental equation* [1, 8] applied to the system and surroundings, gives

$$T dS - Y dX = -T_e dS_e + Y_e dX_e, \tag{A1}$$

where the subscript 'e' (for exterior) denotes the variables of the surroundings. (A1) describes a thermodynamic process as an interaction between the system and its surroundings, which contains all its physics. Thus, heat and work, which are process functions representing energy crossing the boundary that separates the system from its surroundings, cannot add anything essential to (A1).

Taking the Zemansky definition [1], in line with the reservoir concept [9], work is energy that crosses the boundary and whose overall effect can be described as 'the alteration of the position or configuration of some external mechanical device'. This is the ultimate criterion that determines whether the energy crossing the boundary is work or not [1]. The *work* is thus given by the term $Y_e dX_e$ in (A1), i.e.

$$\delta W = Y_e dX_e. \tag{A2}$$

Because each side of (A1) is the internal energy variation of the system, $dU$, which by the first law of thermodynamics is $\delta Q + \delta W$, where $\delta Q$ is the *heat* (energy that crosses the boundary), (A2) implies that $\delta Q$ has to be given by the term $-T_e dS_e$ in (A1), i.e.

$$\delta Q = -T_e dS_e. \tag{A3}$$

The term $-Y dX$ in (A1) is the *configuration work*, $\delta W_C$,

$$\delta W_C = -Y dX, \tag{A4}$$

and the difference $\delta W - \delta W_C$ is the *dissipative work*, $\delta W_D$,

$$\delta W_D = Y_e dX_e + Y dX. \tag{A5}$$

From the previous equations (A2)–(A5), $T dS$ in (A1) and $dU$ can be written as

$$T dS = (\delta Q + \delta W_D), \tag{A6}$$

$$dU = (\delta Q + \delta W_D) + \delta W_C. \tag{A7}$$

The brackets in (A6) and (A7) stress that heat, $\delta Q$, and dissipative work, $\delta W_D$, produce indistinguishable effects on the system. However, they cannot be perceived as the same, because heat changes the entropy of surroundings, given by (A3), but dissipative work does not.

It is instructive to apply this formalism to the problem of determining thermal capacity, presented at the beginning of the paper. The temperature of the system (e.g. a given mass of water) can be raised by heat from a flame (which belongs to the surroundings) or by using an electrical resistance (which belongs to the system) connected to a battery (which belongs to the surroundings). The system is open to the atmosphere at pressure $P_e$, equal to that of the system, $P$. Therefore, there are three energy interactions to consider. The heat from the flame is given by (A3). The electrical interaction is characterised by the electric charge, $q_e$, and the electromotive force, $\varepsilon_e$, of the battery, both surroundings variables; however, for the sake of formal generality, for the system, we also consider an electric charge, $q$, and an electromotive force, $\varepsilon$, which are zero here. Thus, for this interaction, the generalised mechanical variables become:

$$Y = \varepsilon; \; Y_e = \varepsilon_e; \; dX = -dq; \; dX_e = -dq_e. \tag{A8}$$

As energy entering the system is taken as positive, and as negative otherwise, the minus signs in (A8) are needed because the decrease in battery charge corresponds to energy entering the system. Lastly, the interaction resulting from the system expansion, in which the volumes of the system and surroundings undergo symmetric variations, that is, $dV = -dV_e$. In this case, the generalised mechanical variables are:

$$Y = P; \; Y_e = P_e; \; dX = dV; \; dX_e = dV_e. \tag{A9}$$

Using (A3), (A5), (A8) and (A9), for the entire process, (A6) becomes

$$TdS = -T_e dS_e + PdV + P_e dV_e - \varepsilon dq - \varepsilon_e dq_e. \tag{A10}$$

The first term on the right-hand side of (A10) is the heat, $\delta Q$, and the sum of the other four is the dissipative work, $\delta W_D$. However, as in this problem $P = P_e$, $dV = -dV_e$, $\varepsilon = 0$, and $q = 0$, (A10) becomes

$$TdS = -T_e dS_e - \varepsilon_e dq_e. \tag{A11}$$

Using the definition of electric current and Ohm's law, $-dq_e = (\varepsilon_e/r)dt$, where $r$ is the resistance and $dt$ is the infinitesimal heating time, $TdS$ can be rewritten as

$$TdS = -T_e dS_e + (\varepsilon_e^2/r)dt. \tag{A12}$$

The term $\left(\varepsilon_\mathrm{e}^2/r\right)\mathrm{d}t$ in (A12), i.e. the energy transferred from the battery to the system, is work since it can be imagined as energy obtained from a body falling in the gravitational field while driving an electric generator connected to the resistance [1].

## References

[1] Zemansky M W and Dittman R H 1997 *Heat and Thermodynamics* 7th edn (New York: McGraw-Hill)

[2] Anacleto J and Ferreira J M 2018 Why is dissipative work insistently ignored? The case of heat capacities *Eur. J. Phys.* **39** 055102

[3] Anacleto J 2021 Thermal capacities: system or process properties? *Eur. J. Phys.* **42** 025102

[4] Callen H B 1985 *Thermodynamics and an Introduction to Thermostatistics* 2nd edn (New York: John Wiley)

[5] Swendsen R H 2012 *An Introduction to Statistical Mechanics and Thermodynamics* (Oxford: Oxford University Press)

[6] Knight R 2022 All About Polytropic Processes *Phys. Teach.* **60** 422-24

[7] Borgnakke C and Sonntag R E 2009 *Fundamentals of Thermodynamics* 7th edn (New Jersey: John Wiley)

[8] Güémez J, Fiolhais C and Fiolhais M 2000 Equivalence of thermodynamical fundamental equations *Eur. J. Phys.* **21** 395-404

[9] Anacleto J 2021 The reservoir concept: entropy generation and lost work *Eur. J. Phys.* **42** 035102